\documentclass[12pt]{article}

\usepackage{times}
\usepackage{amsfonts}

  \topmargin - 0.5 cm 
\begin{document}

\baselineskip = 0.60 true cm

\begin{center}
{\large \bf Variations on a theme of Heisenberg, Pauli and 
Weyl\footnote{Dedicated to the memory of my teacher and friend 
Mosh\'e Flato on the occasion of the tenth anniversary of his death.}}\\ 
\end{center}

\vspace{0.5cm}

\begin{center}
{\bf Maurice R Kibler}
\end{center}

\begin{center}
{Universit\'e de Lyon, F--69622, Lyon, France; \\
Universit\'e Lyon 1, Villeurbanne; \\
CNRS/IN2P3, UMR5822, Institut de Physique Nucl\'eaire de Lyon} \\
email: m.kibler@ipnl.in2p3.fr
\end{center}

\vspace{1.25cm}



\vspace{1cm}

\noindent {\bf Abstract} 

The parentage between Weyl pairs, generalized Pauli group and unitary 
group is investigated in detail. We start from an abstract definition 
of the Heisenberg-Weyl group on the field $\mathbb{R}$ and then switch 
to the discrete Heisenberg-Weyl group or generalized Pauli group on a 
finite ring $\mathbb{Z}_d$. The main characteristics of the latter group, 
an abstract group of order $d^3$ noted $P_d$, are given (conjugacy classes 
and irreducible representation classes or equivalently Lie algebra of 
dimension $d^3$ associated with $P_d$). Leaving the abstract sector, a 
set of Weyl pairs in dimension $d$ is derived from a polar decomposition 
of $SU(2)$ closely connected to angular momentum theory. Then, a realization 
of the generalized Pauli group $P_d$ and the construction of generalized 
Pauli matrices in dimension $d$ are revisited in terms of Weyl pairs. Finally, 
the Lie algebra of the unitary group $U(d)$ is obtained as a subalgebra of the 
Lie algebra associated with $P_d$. This leads to a development of the Lie 
algebra of $U(d)$ in a basis consisting of $d^2$ generalized Pauli matrices. In 
the case where $d$ is a power of a prime integer, the Lie algebra of $SU(d)$ 
can be decomposed into $d-1$ Cartan subalgebras.

   \newpage

\section{Introduction}

The present paper is devoted to three major ingredients of quantum mechanics, 
namely, the Heisenberg-Weyl group connected with Heisenberg commutation 
relations \cite{Heisenberg}, the Pauli spin matrices \cite{Pauli} used in 
generalized angular momentum theory and theory of unitary groups, and 
the pairs of Weyl \cite{Weyl} of relevance in finite quantum mechanics.

The Heisenberg-Weyl (or Weyl-Heisenberg or Heisenberg) group $HW(\mathbb{R})$, also called the 
Weyl group \cite{Wolf1}, is of central importance for the quantization process 
and its Lie algebra turns out to be a basic building unit for quantum mechanics 
\cite{Wolf2}. Note that the Lie algebra of $HW(\mathbb{R})$ should not be 
confused with the Weyl-Heisenberg algebra (or oscillator algebra spanned by the 
creation, annihilation and number operators) and its supersymmetric extensions 
$W_k$ \cite{DaoKib}. 

A discrete restriction $HW(\mathbb{Z}_d)$ of $HW(\mathbb{R})$, corresponding to 
the replacement of the infinite field $\mathbb{R}$ by a finite ring 
$\mathbb{Z}_d \equiv \mathbb{Z}/d\mathbb{Z}$, yields a group of order $d^3$ 
($d$ arbitrary in $\mathbb{N} \setminus \{ 0,1 \}$). This group was introduced 
by \v{S}\v{t}ov\'{\i}\v{c}ek and Tolar \cite{Tolar1} in connection with quantum 
mechanics in a discrete space-time, by Balian and Itzykson in connection with 
finite quantum mechanics \cite{Balian}, and by Patera and Zassenhaus  
\cite{PateraZassenhaus} in connection with gradings of simple Lie algebras of 
type $A_{n-1}$. The case where the ring $\mathbb{Z}_d$ is replaced by a finite 
(Galois) field $\mathbb{F}_q$ gave rise to several mathematical studies 
\cite{TerrasHowe, Howe}. The discrete Heisenberg-Weyl group, also known as the 
generalized Pauli group, plays a central role in quantum information, 
cf. the interest of Galois fields in finite quantum mechanics \cite{Vourdas1}
and, consequently, in quantum information and quantum computation. In this 
connection, a finite Heisenberg-Weyl group was used for a description of phase 
oscillations of EPR states \cite{Ajout}.
 
What is the relationship between the Heisenberg-Weyl group and Weyl pairs? 
First of all, a definition of a Weyl pair is in order. A Weyl pair $(X , Z)$ 
in $d$ dimensions is a pair of $d$-dimensional unitary matrices $X$ and $Z$ 
that satisfy the $q$-commutation relation $XZ - qZX = 0$ and the cyclic relations 
$X^d = Z^d = I$ ($I$ standing here for the unitary matrix), where $q$ is a 
primitive root of unity with $q^d = 1$. The concept 
of a pair of Weyl, initially introduced for dealing with quantum dynamical 
systems in finite dimension \cite{Weyl}, was used for the construction 
of unitary bases in finite-dimensional Hilbert spaces \cite{Schwinger} and 
(independently) for the factorization of the secular equation corresponding to 
finite-dimensional eigenvalue problems \cite{McIntosh}. In the last 20 years, the 
notion of Weyl pairs was used for the construction of {\it generalized} Pauli 
matrices in domains as different as graded Lie algebras and quantum information.

The {\it usual} Pauli matrices $\sigma_x$, $\sigma_y$ and $\sigma_z$ are useful 
for the representation theory of the Lie group $SU(2)$. Therefore, a natural 
extension of the Pauli matrices resulted in the sixties from the interest of the group $SU(3)$ 
for the classification of elementary particles \cite{Gell-MannNe'emanOkubo}. 
This gave gise to the Gell-Mann matrices and the Okubo matrices. Further 
extensions of the Pauli matrices came out of the introduction of the group 
$SU(4)$ for charmed particle \cite{GIM} and of the group $SU(5)$ for a grand unified 
theory of quarks and leptons \cite{Georgi-Glashow}. The Gell-Mann lambda 
matrices for $SU(3)$ and their extension to Cartan bases for $SU(d)$ 
undoubtedly constitute a systematic extension of the ordinary Pauli matrices. 
This statment is particularly justified as far as the tensor structure (involving 
symmetric and antisymmetric tensors) of their algebra is concerned \cite{Azcarraga}. 
We shall deal in this paper with another extension of the Pauli matrices in $d$ 
dimensions which turns out to be of special interest in the case where
$d$ is a power of a prime integer. Indeed, generalized Pauli matrices can 
be constructed in a systematic way by making use of Weyl pairs. In this 
direction let us mention the pioneer work of Patera and Zassenhaus 
\cite{PateraZassenhaus}. In the last two decades, the construction of 
generalized Pauli spin matrices has been extensively used 
in the theory of semi-simple Lie algebras, in quantum mechanics 
(complete state determination, reconstruction of a density matrix, discrete 
Wigner functions), in quantum information and quantum computation 
(mutually unbiased bases, unitary error bases, quantum error correction, 
random unitary channels, mean king's problem, positive operator valued measures,
and quantum entanglement), and in the study of modified Bessel functions (see 
for instance \cite{Balian, PateraZassenhaus} and \cite{KnillGPM1}-\cite{Kazuyuki}). 

From a group-theoretical point of view, the $d$-dimensional generalized Pauli 
matrices may serve to construct a generalized Pauli group in $d$ dimensions,  
a group generalizing the ordinary Pauli group spanned by the ordinary Pauli 
matrices (see \cite{Tolar1}-\cite{Vourdas1}, \cite{GottesmanGPM2}, 
\cite{BandyoGPM5} and \cite{Lawrence2}-\cite{PlaJor}). In fact, this group is 
nothing but the discrete Heisenberg-Weyl group $HW(\mathbb{Z}_d)$. This 
generalized Pauli group has been recently the object of numerous studies
partly in connection with the Clifford or Jacobi group 
\cite{Grassl, Appleby, Flammia, Cormick} as well as graph-theoretical 
and finite-geometrical analyses of the generalized Pauli operators 
\cite{geometrical analysis of Pd, PlaJor} .

The object of this work is to further study the link between the 
Heisenberg-Weyl group, the Weyl pairs, the generalized Pauli matrices and the 
generalized Pauli group and to revisit their interest for unitary 
groups. We shall start with an abstract definition of the Heisenberg-Weyl 
group, pass to an abstract version of $HW(\mathbb{Z}_d)$ and briefly study 
it. Then, we shall deal with the introduction of Weyl pairs 
from a polar decomposition of the Lie algebra $su(2)$ and we shall use them for 
finding a realization of $HW(\mathbb{Z}_d)$ isomorphic to the generalized 
Pauli group in $d$ dimensions. Finally, some of the generators of the Pauli 
group in $d$ dimensions shall be used for constructing the Lie algebra 
$su(d)$ of $SU(d)$ in a basis that is especially adapted, when $d$ is a 
power of a prime integer, to a decomposition of $su(d)$ into a direct sum of 
$d+1$ Cartan subalgebras. 

\section{The Heisenberg-Weyl group}

\subsection{The Lie group $HW(\mathbb{R})$}
We start with an abstract definition of the Heisenberg-Weyl group $HW(\mathbb{R})$. Let
us consider the set of triplets
          \begin{eqnarray}
S := \{ (x, y, z) : x, y, z \in \mathbb{R} \}.           
          \end{eqnarray} 
The set $S$ can be equipped with the internal composition law $S \times S \to S$ defined trough 
          \begin{eqnarray}
(x, y, z) (x', y', z') := (x + x' - zy', y + y', z + z'). 
          \label{law HW(R)} 
          \end{eqnarray} 
It is clear that the set $S$ is a group with respect to the law (\ref{law HW(R)}). We denote 
$HW(\mathbb{R})$ this group and call it the Heisenberg-Weyl group (for evident reasons to be 
given below) on the infinite field $\mathbb{R}$. More precisely, we have the following result.

\noindent {\it Proposition 1}. The group $HW(\mathbb{R})$ is a noncompact Lie group of order 
3. This nonabelian group is nilpotent (hence solvable) with a nilpotency class equal to 2. 

\noindent {\it Proof}. The proof is trivial. Let us simply mention that the 
nilpotency of $HW(\mathbb{R})$ follows by repeated use of the commutator 
          \begin{eqnarray}
          [(x', y', z') , (x, y, z)] = (zy' - yz', 0, 0) 
          \label{commutator HW(R)} 
          \end{eqnarray} 
of the elements $(x', y', z')$ and $(x, y, z)$ of the group $HW(\mathbb{R})$. Equation 
(\ref{commutator HW(R)}) shows that 
$(x, y, z)$ and $(x', y', z')$ commute if and only if $zy' - yz' = 0$. $\Box$

In the terminology of Wigner \cite{Wigner}, the group $HW(\mathbb{R})$ 
is not ambivalent (ambivalent means that each conjugacy class contains 
its inverse elements). Indeed, since
          \begin{eqnarray}
(x, y, z)^{-1} = (-x - yz, -y, -z) 
          \end{eqnarray} 
and           
          \begin{eqnarray}
(x', y', z') (x, y, z) (x', y', z')^{-1} = (x + zy' - yz', y, z)
          \end{eqnarray} 
it is evident that only the class ${\cal C}_{(0, 0, 0)} = \{ (0, 0, 0) \}$ 
of the identity element $(0, 0, 0)$ is ambivalent.

\subsection{The Lie algebra of $HW(\mathbb{R})$}
We may ask why to call $HW(\mathbb{R})$ the Heisenberg-Weyl group? The following result clarifies 
this point. 

\noindent {\it Proposition 2}. A set of infinitesimal generators of $HW(\mathbb{R})$ is
          \begin{eqnarray}
{\cal H} = \frac{1}{i}        \frac{\partial}{\partial x} \qquad 
{\cal Q} = \frac{1}{i}        \frac{\partial}{\partial y} \qquad
{\cal P} = \frac{1}{i} \left( \frac{\partial}{\partial z} - y \frac{\partial}{\partial x} \right).
          \end{eqnarray} 
This set of generators satisfies the formal commutation relations
          \begin{eqnarray}
[Q , P]_- = i H \qquad 
[P , H]_- = 0   \qquad 
[H , Q]_- = 0
          \label{Lie brackets} 
          \end{eqnarray} 
with $H = {\cal H}$, $Q = {\cal Q}$ and $P = {\cal P}$. The Lie algebra $hw(\mathbb{R})$ 
of $HW(\mathbb{R})$, with the Lie brackets (\ref{Lie brackets}), is a three-dimensional 
nilpotent (hence solvable) Lie algebra with nilpotency class 2.

\noindent {\it Proof}. The proof easily follows by working in a neighbourhood of the identity 
$(0, 0, 0)$ of $HW(\mathbb{R})$ and by considering the series $w^1 = hw(\mathbb{R})$, 
$w^2 = [w^1 ,w^1]_-$, $w^3 = [w^1 ,w^2]_-, \cdots$ where $[A , B]_-$ refers here to the set 
of commutators $[\alpha , \beta]_-$ with $\alpha \in A$ and $\beta \in B$. $\Box$

The connection with the Heisenberg commutation 
relations is clearly emphasized by (\ref{Lie brackets}). This constitutes a partial justification 
for calling $HW(\mathbb{R})$ the Heisenberg-Weyl group on $\mathbb{R}$. The Lie algebra 
$hw(\mathbb{R})$ was derived from a matrix group \cite{Wolf1} and studied at length from the point 
of view of quantum mechanics \cite{Wolf2}. This algebra admits infinite-dimensional  
representations by Hermitean matrices. In particular, we have the infinite-dimensional harmonic 
oscillator representation which is associated with the operator realization 
$H = {\cal H}_{ho} := {\hbar} 1$, 
$Q = {\cal Q}_{ho} :=  x$ and 
$P = {\cal P}_{ho} :=  \frac{\hbar}{i} \frac{\partial}{\partial x}$, where 
${\hbar}$ is the rationalized Planck constant. On the other side, we may expect to have finite-dimensional 
representations of $hw(\mathbb{R})$ at the price to abandon the Hermitean character of the representation 
matrices. 

As an example, we have the three-dimensional representation of $hw(\mathbb{R})$ 
defined by $H = H_{3}$, $Q = Q_{3}$ and $P = P_{3}$ with 
          \begin{eqnarray}
H_{3} :=  
\pmatrix{
0      &    0 &      0   \cr
0      &    0 &      0   \cr
i      &    0 &      0   \cr
}
\qquad
Q_{3} :=  
\pmatrix{
0      &    0 &      0   \cr
i      &    0 &      0   \cr
0      &    0 &      0   \cr
}
\qquad
P_{3} :=  
\pmatrix{
0      &    0  &      0   \cr
0      &    0  &      0   \cr
0      &    -i &      0   \cr
}.
          \label{rep 3x3} 
          \end{eqnarray}
We can look for the matrix Lie group which corresponds to the Lie algebra spanned by the set 
$\{ H_{3}, Q_{3}, P_{3} \}$. This yields Proposition 3.

\noindent {\it Proposition 3}. The exponentiation
          \begin{eqnarray}
M(x, y, z) := \exp [i (x H_{3} + y Q_{3} + z P_{3} )]
          \label{exponentiation} 
          \end{eqnarray} 
leads to 
          \begin{eqnarray}
M(x, y, z) = 
\pmatrix{
1                   &    0  &      0   \cr
-y                  &    1  &      0   \cr
-x -\frac{1}{2}yz   &    z  &      1   \cr
}.
          \label{matrix M} 
          \end{eqnarray} 
The matrices $M(x, y, z)$ satisfy the composition law
          \begin{eqnarray}
M(x, y, z) M(x', y', z') =  M(x + x' + \frac{1}{2} zy' - \frac{1}{2} yz' , y + y', z + z') 
          \label{loi en M} 
          \end{eqnarray} 
so that the set $S' := \{ M(x, y, z) : x, y, z \in \mathbb{R} \}$ 
endowed with the law (\ref{loi en M}) is a group isomorphic to $HW(\mathbb{R})$.  

\noindent {\it Proof}. A simple expansion of (\ref{exponentiation}) where $H_{3}$, $Q_{3}$ and $P_{3}$
are given by (\ref{rep 3x3}) yields (\ref{matrix M}). The isomorphism follows from the bijection 
$S \to S' : (x,y,z) \mapsto M(- x - \frac{1}{2}yz, -y, -z)$. 
Note that the matrix form (\ref{matrix M}) corresponds 
to two other sets $\{ {\cal H}_{\pm}, {\cal Q}_{\pm}, {\cal P}_{\pm} \}$ 
of infinitesimal generators of $HW(\mathbb{R})$, namely, 
          \begin{eqnarray}
{\cal H}_{\pm} = \pm i        \frac{\partial}{\partial x} \qquad 
{\cal Q}_{\pm} = \pm i \left( \frac{\partial}{\partial y} \mp  \frac{1}{2}  z \frac{\partial}{\partial x} \right) \qquad
{\cal P}_{\pm} = \pm i \left( \frac{\partial}{\partial z} \pm  \frac{1}{2}  y \frac{\partial}{\partial x} \right)
          \end{eqnarray} 
which satisfies (\ref{Lie brackets}) with 
$H = {\cal H}_{\pm}$, 
$Q = {\cal Q}_{\pm}$ and 
$P = {\cal P}_{\pm}$ (cf. \cite{Wolf1, Wolf2}). $\Box$
 
\section{The Pauli group}

\subsection{The abstract Pauli group}
\subsubsection{The group $P_d$}
We shall be concerned in this section with a discretization of the 
Heisenberg-Weyl group $HW(\mathbb{R})$. A trivial discretization of 
$HW(\mathbb{R})$ can be obtained by replacing the field $\mathbb{R}$ by the 
infinite ring $\mathbb{Z}$. This leads to an infinite-dimensional discrete 
group $HW(\mathbb{Z})$. A further possibility is to replace $\mathbb{R}$ by 
the finite ring 
$\mathbb{Z}_d \equiv \mathbb{Z}/d\mathbb{Z}$ where $d$ is arbitrary in 
$\mathbb{N} \setminus \{ 0,1 \}$. (In the case where $d$ is a prime $p$ or 
a power of a prime $p^e$ with $e \in \mathbb{N} \setminus \{ 0,1 \}$, the 
finite ring $\mathbb{Z}/d\mathbb{Z}$ can be replaced by the Galois field 
$\mathbb{F}_p$ or $\mathbb{F}_{p^e}$.) This yields a finite group 
$HW(\mathbb{Z}_d)$ which can be described by the following result. 

\noindent {\it Proposition 4}. The set 
          \begin{eqnarray}
S_d := \{ (a, b, c) : a, b, c \in \mathbb{Z}_d \}           
          \end{eqnarray} 
with the internal composition law $S_d \times S_d \to S_d$ defined trough 
          \begin{eqnarray}
(a, b, c) (a', b', c') := (a + a' - cb', b + b', c + c')
          \label{law Pd} 
          \end{eqnarray} 
(where from now on the addition is understood modulo $d$) is a finite group of order $d^3$. This nonabelian 
group $HW(\mathbb{Z}_d)$, noted $P_d$ for short, is nilpotent (hence solvable) with 
a nilpotency class equal to 2. 

\noindent {\it Proof}. The proof of Proposition 4 is elementary. Note simply 
that we have the canonical decomposition  
          \begin{eqnarray}
(a, b, c) = (a, 0, 0) (0, b, 0) (0, 0, c) 
          \label{Euler decomposition}
          \end{eqnarray} 
for any element $(a, b, c)$ of $P_d$ and that two elements $(a, b, c)$ and $(a', b', c')$ of $P_d$ commute 
if and only if $cb' - bc' = 0$ (mod $d$). $\Box$

We call the abstract group $P_d$ the {\it (generalized) Pauli group} in $d$ dimensions. At this stage, we can give the 
main reason for associating Heisenberg, Pauli and Weyl in the title of the present paper. As a point 
of fact, the discretization of the group $HW(\mathbb{R})$, a group associated with the {\it Heisenberg 
commutation relations}, via the replacement $\mathbb{R} \to \mathbb{Z}/d\mathbb{Z}$ gives rise to 
the group $P_d$, a group which can be realized in terms of {\it generalized Pauli matrices}, which in turn 
can be constructed in terms of {\it Weyl pairs} (see below).    

\subsubsection{Some subgroups of $P_d$}
Among the subgroups of $P_d$, we can mention proper subgroups of order $d$ and $d^2$ 
(there are no other proper subgroups if $d$ is a prime integer). We simply list below 
the subsets of $S_d$, which together with the law (\ref{law Pd}), provide us with some important 
subgroups of $P_d$. 

- The set $\{ (a, 0, 0) : a \in \mathbb{Z}_d \}$ gives an invariant abelian subgroup of 
$P_d$ of order $d$ isomorphic to the cyclic group $\mathbb{Z}_d$. In fact, this subgroup is the centrum 
$Z(P_d)$ of $P_d$ and $P_d/Z(P_d)$ is isomorphic to $\mathbb{Z}_d \otimes \mathbb{Z}_d$. 

- The set $\{ (0, b, 0) : b \in \mathbb{Z}_d \}$ gives an abelian subgroup of 
$P_d$ of order $d$ isomorphic to $\mathbb{Z}_d$. 

- Similarly, the set $\{ (0, 0, c) : c \in \mathbb{Z}_d \}$ gives also an abelian subgroup of 
$P_d$ of order $d$ isomorphic to $\mathbb{Z}_d$. 

- The sets $\{ (a, b, 0) : a, b \in \mathbb{Z}_d \}$ 
       and $\{ (a, 0, c) : a, c \in \mathbb{Z}_d \}$ 
give two invariant abelian subgroups of $P_d$ of order $d^2$ isomorphic to $\mathbb{Z}_d \otimes \mathbb{Z}_d$. 

- Finally, the set $\{ (a, b, b) : a, b \in \mathbb{Z}_d \}$ give an invariant 
abelian subgroup of $P_d$ of order $d^2$. 

\subsubsection{Conjugacy classes of $P_d$}
The conjugacy classes of $P_d$ readily follow from 
          \begin{eqnarray}
(a', b', c') (a, b, c) (a', b', c')^{-1} = (a + cb' - bc', b, c)  
          \end{eqnarray} 
with addition modulo $d$. This can be precised by the following result. 

\noindent {\it Proposition 5}. The group $P_d$ has $d(d+1)-1$ conjugacy classes: $d$ classes containing each 
1 element and $d^2 - 1$ classes containing each $d$ elements.

\noindent {\it Proof}. It can be checked that the class 
${\cal C}_{(a, 0, 0)}$ of $(a, 0, 0)$ is 
${\cal C}_{(a, 0, 0)} = \{ (a, 0, 0) \}$; therefore, there are $d$ 
classes with 1 element. Furthermore, the class 
${\cal C}_{(a, b, c)}$ of $(a, b, c)$, with the case $b=c=0$ excluded, is 
${\cal C}_{(a, b, c)} = \{ (a', b, c) : a' \in \mathbb{Z}_d \} $; this yields $d^2 - 1$ classes with 
$d$ elements. We note that the group $P_d$ is not ambivalent in general. $\Box$

The case $d=2$ is very special since the group $P_2$ of order 8 is ambivalent like the group $Q$ of ordinary quaternions, 
another group of order 8. Not all the subgroups of $P_2$ are invariant. Therefore, the group $P_2$ is not 
isomorphic to $Q$ (for which all subgroups are invariant). Indeed, it can be proved that $P_2$ is 
isomorphic to the group of hyperbolic quaternions associated with the Cayley-Dickson algebra $A(c_1 , c_2)$ 
with $(c_1, c_2) \not= (-1, -1)$ defined in \cite{Lambert}. In this respect, the Pauli group $P_2$ 
defined in this work differs from the Pauli group in $d=2$ dimensions considered by some authors, 
a group isomorphic to the group $Q$ of ordinary quaternions. Let $P_2'$ be this latter Pauli group. 
It consists of the elements 
$\sigma := \pm \sigma_0$, $\pm i \sigma_x$, $\pm i \sigma_y$, $\pm i \sigma_z$ 
(where $\sigma_0$ is the $2 \times 2$ unit matrix). Let us also mention that an extension of the 
group $P_2'$ is used in quantum computation \cite{Nielsen} (see also \cite{Lawrence2, PlaJor}). This extension, 
say $P_2''$, is obtained from a doubling process: The group $P_2''$ consists of the elements of the 
set $\{ \sigma, i \sigma : \sigma \in P_2' \}$. Thus, the conjugation classes and the irreducible 
representation classes of $P_2''$ trivially follow from those of $P_2'$. 

\subsubsection{Irreducible representations of $P_d$}
The duality between conjugacy classes and classses of irreducible representations 
leads to the following result. 

\noindent {\it Proposition 6}. The group $P_d$ has $d(d+1) - 1$ classes of irreducible 
representations: $d^2$ classes of dimension 1 and $d - 1$ classes of dimension $d$.

\noindent {\it Proof}. It is sufficient to apply the Burnside-Wedderburn theorem. $\Box$

As a corollary of Proposition 5 and Proposition 6, the difference between the 
order of $P_d$ and its number of classes (conjugacy classes or irreducible 
representation classes) is odd if $d=2k$ ($k \in \mathbb{N}^*$), or a multiple 
of 16 if $d = 4k+3$ ($k \in \mathbb{N}$) or a multiple 
of 32 if $d = 4k+1$ ($k \in \mathbb{N}^*$). (For an arbitrary finite group of 
odd order, the difference is a multiple of $16$.) Furthermore, the number of 
elements of $P_d$ which commute with a given element $(a, b, c)$ of $P_d$ is 
$d^3$ or a multiple of $d^2$ according to whether the order of the conjugation 
class containing $(a, b, c)$ is $1$ or $d$; see \cite{geometrical analysis of Pd} 
for a more elaborated result, in the form of a universal formula, and its 
interpretation in terms of the fine structure of the projective line defined 
over the modular ring $\mathbb{Z}_d$. Note that Proposition 5 and Proposition 
6 are in agreement with the results obtained \cite{TerrasHowe} in the case where 
$d$ is a power of a prime integer corresponding to the replacement of the ring 
$\mathbb{Z}_d$ by the Galois field $\mathbb{F}_d$.

\subsubsection{A Lie algebra associated with $P_d$}
We close the study of the abstract group $P_d$ with a result devoted to the association of $P_d$ 
with a Lie algebra of dimension $d^3$. Let us consider the group algebra (or Frobenius algebra) 
$F(P_d)$ of the generalized Pauli group $P_d$. Such an algebra is an associative algebra over 
the field $\mathbb{C}$. By applying the process developed by Gamba \cite{Gamba}, we can 
construct from $F(P_d)$ a Lie algebra, which we shall denote as $p_d$, by taking 
\begin{eqnarray}
\langle (a,b,c) , (a',b',c') \rangle := (a+a'-cb', b+b', c+c') - (a+a'-bc', b+b', c+c') 
\label{Lie brackets of pd}
\end{eqnarray}
for the Lie bracket of $(a,b,c)$ and $(a',b',c')$. (The right-hand side of (\ref{Lie brackets of pd}) 
is defined in $F(P_d)$.) The set $S_d$ constitutes a basis both for the Frobenius algebra $F(P_d)$ and 
the Lie algebra $p_d$ ($S_d$ is a Chevalley basis for $p_d$). As a further result, we have the 
following proposition. 

\noindent {\it Proposition 7}. The Lie algebra $p_d$ of dimension $d^3$, associated with the 
finite group $P_d$ of order $d^3$, is not-semi-simple. It can be decomposed as the direct sum
\begin{eqnarray}
p_d = \bigoplus_{1}^{d^2} {u}(1) \bigoplus_{1}^{d-1} {u}(d)  
\label{decomposition of pd} 
\end{eqnarray}
which contains $d^2$ Lie algebras isomorphic to $u(1)$ and $d-1$ 
Lie algebras isomorphic to $u(d)$. 

\noindent {\it Proof}. The proof can be achieved by passing from the 
Chevalley basis of $p_d$, inherent to (\ref{Lie brackets of pd}), to 
the basis generated by the idempotent (or projection) operators and 
nilpotent (or ladder) operators, defined in $F(P_d)$, associated with 
the classes of irreducible representations of $P_d$. Equation 
(\ref{decomposition of pd}) is reminiscent of the fact that $P_d$ has $d^2$ 
irreducible representation classes of dimension $1$ and $d-1$ irreducible 
representation classes of dimension $d$. $\Box$  

\subsection{A realization of the Pauli group}
\subsubsection{Polar decomposition of $SU(2)$}

Let ${\cal E} (2j+1)$, with $2j \in \mathbb{N}$, be a $(2j + 1)$-dimensional 
Hilbert space of constant angular momentum $j$. Such a space is spanned by the set 
$\{ |j , m \rangle : m = j, j-1, \cdots, -j \}$, where $|j , m \rangle$ is an
eigenstate of the square $j^2$ and the $z$-component $j_z$ of a generalized
angular momentum \cite{Edmonds}. The state vectors $|j , m \rangle$ are taken 
in an orthonormalized form, i.e., the inner product 
$\langle j , m | j' , m' \rangle$ is equal to $\delta_{m,m'}$. 

Following the approach of \cite{Kib polar decomp}, 
we define the linear operator $v_{ra}$ via 
          \begin{eqnarray}
  v_{ra} |j , m \rangle = \left( 1 - \delta_{m,j} \right) q^{(j-m)a}
  |j , m+1 \rangle + \delta_{m,j}  
  {e}^{{i} 2 \pi j r}
  |j , -j \rangle 
          \label{action de vra sur jm} 
          \end{eqnarray}
where
	  \begin{eqnarray}
     r \in \mathbb{R} \qquad 
     a \in \mathbb{R} \qquad 
     q = \exp \left( {2 \pi {i} \over 2j+1} \right). 
        \label{definition of q}
        \end{eqnarray}
The matrix $V_{ra}$ of the operator $v_{ra}$ in the spherical basis 
    \begin{eqnarray}
b_s : = \{ |j , j \rangle, |j , j-1 \rangle, \cdots, |j , -j \rangle \}  
    \label{spherical basis}
    \end{eqnarray} 
reads
        \begin{eqnarray}
V_{ra} = 
\pmatrix{
0                    &    q^a &      0  & \cdots &       0 \cr
0                    &      0 & q^{2a}  & \cdots &       0 \cr
\vdots               & \vdots & \vdots  & \cdots &  \vdots \cr
0                    &      0 &      0  & \cdots & q^{2ja} \cr
{e}^{{i} 2 \pi j r}  &      0 &      0  & \cdots &   0     \cr
}.
        \label{definition of Vra}
        \end{eqnarray}
The matrix $V_{ra}$ constitutes a generalization of the matrix $V_a$ 
introduced in \cite{KiblerPlanat} (see also \cite{AlbouyKibler}). 

The shift operator $v_{ra}$ takes its origin in the study of the Lie algebra of
$SU(2)$ in a nonstandard basis with the help of two quon algebras describing
$q$-deformed oscillators \cite{quon algebras}.
The operator $v_{ra}$ is unitary. Furthermore, it is cyclic 
in the sense that
          \begin{eqnarray}
  \left( v_{ra} \right)^{2j+1} =  {e} ^{ {i} 2{\pi} j (a + r) } I 
          \label{cyclic}
          \end{eqnarray}  
where $I$ is the identity. The eigenvalues and eigenvectors of $v_{ra}$ 
are given by the following result.

\noindent {\it Proposition 8}. The spectrum of the operator $v_{ra}$ 
is nondegenerate. For fixed $j$, $r$ and $a$, it follows from
          \begin{eqnarray}
  v_{ra} | j \alpha ; r a \rangle  =  q^{j(a+r) - \alpha} 
      | j \alpha ; r a \rangle 
          \label{eigenvalues}
          \end{eqnarray}
where 
          \begin{eqnarray}
|j \alpha ; r a \rangle = \frac{1}{\sqrt{2j+1}} \sum_{m = -j}^{j} 
q^{(j + m)(j - m + 1)a / 2 - j m r + (j + m)\alpha} | j , m \rangle
          \label{j alpha r a in terms of jm}
          \end{eqnarray} 
for $\alpha = 0, 1, \cdots, 2j$.

A second linear operator is necessary to define a polar decomposition of 
$SU(2)$. Let us introduce the Hermitean operator $h$ through 
        \begin{eqnarray}
   h |j , m \rangle = {\sqrt{ (j+m)(j-m+1) }} |j , m \rangle. 
        \end{eqnarray}
Then, it is a simple matter of calculation to show that the three operators 
	  \begin{eqnarray}
  j_+ = h           v_{ra}  \qquad  
  j_- = v_{ra}^{\dagger} h  \qquad 
  j_z = \frac{1}{2} ( h^2 - v_{ra}^{\dagger} h^2 v_{ra} )
          \end{eqnarray}
satisfy the ladder equations
     \begin{eqnarray}
  j_+ |j , m \rangle   & = & q^{+(j - m + s - 1/2)a}
  {\sqrt{ (j - m)(j + m+1) }} 
  |j , m + 1 \rangle 
  \\
  j_- |j , m \rangle   & = & q^{-(j - m + s + 1/2)a}
  {\sqrt{ (j + m)(j - m+1) }} 
  |j , m - 1 \rangle 
     \end{eqnarray} 
and the eigenvalue equation 
               \begin{eqnarray} 
  j_z   |j , m \rangle = m |j,m \rangle
               \end{eqnarray}
where $s = 1/2$. (Note that there is one misprint in the corresponding relations of 
\cite{AlbouyKibler}.) Therefore, we have the following result. 

\noindent {\it Proposition 9}. The operators $j_+$, $j_-$ and $j_z$ satisfy the commutation 
relations
     \begin{eqnarray}
  \left[ j_z,j_{+} \right] = + j_{+}  \qquad 
  \left[ j_z,j_{-} \right] = - j_{-}  \qquad 
  \left[ j_+,j_- \right] = 2j_z 
     \label{adL su2}
     \end{eqnarray}
and thus span the Lie algebra of $SU(2)$ over the complex field.

The latter result does not depend on the parameters $r$ and $a$. However, the action 
of $j_+$ and $j_-$ on $|j , m \rangle$ on the space ${\cal E} (2j+1)$ depends on $a$ 
(an {\it a priori} real parameter to be restricted to integer values in what
follows); the usual Condon and Shortley phase convention used in spectroscopy
corresponds to $a = 0$. The writing of the ladder operators $j_+$ and $j_-$ in
terms of $h$ and $v_{ra}$ constitutes a two-parameter polar decomposition of 
the Lie algebra of $SU(1,1)$ [or $SU(2)$over the complex field]. This decomposition 
is an alternative to the polar decompositions obtained independently in 
\cite{other polar decomp, VourdasRPP}.

\subsubsection{Weyl pairs}
The linear operator $x := v_{00}$ such that (cf. (\ref{action de vra sur jm})) 
          \begin{eqnarray}
  x |j , m \rangle = \left( 1 - \delta_{m,j} \right) 
  |j , m+1 \rangle + \delta_{m,j}  
  |j , -j \rangle 
          \label{action de v00 sur jm} 
          \end{eqnarray}
has the spectrum $(1, q, \cdots, q^{2j})$ on ${\cal E}(2j+1)$. Therefore, 
the matrix $X := V_{00}$ of $x$ on the basis $b_s$ is unitarily equivalent to 
          \begin{eqnarray}
Z : = {\rm diag}(1, q, \cdots, q^{2j}).
          \label{definition of Z} 
          \end{eqnarray}
The linear operator $z$ corresponding to the matrix $Z$ can be defined by 
          \begin{eqnarray}
z | j,m \rangle = q^{j-m} | j,m \rangle.
          \label{action de z sur jm} 
          \end{eqnarray}
The two isospectral operators $x$ (a cyclic shift operator) and $z$ 
(a cyclic phase operator) are unitary and constitute a pair of Weyl 
$(x , z)$ since they obey the $q$-commutation relation  
          \begin{eqnarray}
x z - q z x = 0
          \label{q commutation} 
          \end{eqnarray}
(or $X Z - q Z X = 0$ in matrix form). These two operators are 
connected via 
          \begin{eqnarray}
x = f^{\dagger} z f \Leftrightarrow z = f x f^{\dagger}
          \label{x-z connexion} 
          \end{eqnarray}
where $f$ is the Fourier operator such that 
          \begin{eqnarray}
f | j , m  \rangle = \frac{1}{\sqrt{2j+1}} \sum_{m' = -j}^{j} q^{-(j - m)(j - m')} 
  | j , m' \rangle. 
          \end{eqnarray} 
The operator $f$ is unitary and satisfies 
          \begin{eqnarray}
f^4 = 1 
          \end{eqnarray} 
(see \cite{VourdasRPP} for a general treatment of Fourier operators on 
finite-dimensional Hilbert spaces). Let $F$ be the 
matrix of the linear operator $f$ in the basis $b_s$. 
Indeed, $F$ is a circulant matrix. Note that the reduction of the endomorphism 
associated with the matrix $X$ yields the matrix $Z$. In other words, the 
diagonalization of $X$ can be achieved with the help of the matrix $F$ via 
$Z = F X F^{\dagger}$. 

We conclude that the polar 
decomposition of $SU(2)$ described in Section 3.2.1 provides us with 
an alternative derivation of the Weyl pair $(X , Z)$. Of course, other 
pairs of Weyl $(V_{ra}, Z)$, corresponding to $(v_{ra}, z)$ with the 
property $v_{ra} z - q z v_{ra} = 0$, can be derived by replacing 
$v_{00}$ by $v_{ra}$. Note that $v_{ra} =  v_{r0} z^a$.
 
\subsubsection{Weyl pairs and Pauli group}
Let us define the $d^3$ operators 
          \begin{eqnarray}
w_{abc} := q^a x^b z^c \qquad a, b, c \in \mathbb{Z}_d. 
          \end{eqnarray} 
The action of $w_{abc}$ on the Hilbert space ${\cal E} (2j+1)$ is described by 
     \begin{eqnarray}
w_{abc} |j , m \rangle = q^{a + (j-m)c} |j , m + b \rangle 
     \label{action de wabc sur jm} 
     \end{eqnarray}
where $m+b$ is understood modulo $2j+1$. The operators $w_{abc}$ are unitary and satisfy 
          \begin{eqnarray}
{\rm Tr}_{{\cal E}(2j+1)} \left( w_{abc}^{\dagger} w_{a'b'c'} \right) = 
 q^{a' - a} \> d \>
 \delta_{b,b'} \> 
 \delta_{c,c'}
          \label{trace des w(abc)}
          \end{eqnarray} 
with $d := 2j+1$. In addition, we have the following central result. 
 
\noindent {\it Proposition 10}. The set $W_d := \{ w_{abc} : a, b, c \in \mathbb{Z}_d \}$ 
endowed with the multiplication of operators is a group isomorphic to the Pauli group 
$P_d$. Thus, the group $P_d$ is isomorphic to a subgroup of $U(d)$ for $d$ even or 
$SU(d)$ for $d$ odd. 

\noindent {\it Proof}. The proof is immediate: It is sufficient to consider the 
bijection $W_d \to S_d : w_{abc} \mapsto (a,b,c)$, to use repeatedly (\ref{q commutation}) or 
(\ref{action de wabc sur jm}), and to note that the matrix of 
$w_{abc}$ in the basis $b_s$ belongs to $U(d)$ for $d$ even and to $SU(d)$ for $d$ odd. As a 
consequence, the 
Lie bracket $\langle (a,b,c) , (a',b',c') \rangle$, see (\ref{Lie brackets of pd}), 
corresponds to the commutator $[ w_{abc} , w_{a'b'c'} ]_-$ so that the 
Lie algebra $p_d$ associated with the finite group $P_d$ corresponds to 
the commutation relations 
     \begin{eqnarray}
[ w_{a b c} , w_{a' b' c'} ]_- = w_{\alpha \beta \gamma} - w_{\alpha' \beta' \gamma'} 
     \label{commutateur des w(abc)}
     \end{eqnarray}
with $\alpha = a+a'-cb'$, $\beta = b+b'$, $\gamma = c+c'$, 
     $\alpha' = \alpha + cb' - bc'$, $\beta' = \beta$ and $\gamma' = \gamma$. $\Box$

\subsubsection{Weyl pairs and infinite-dimensional Lie algebra}

We close this section by mentioning another interest of Weyl pairs 
$(v_{ra} , z)$. By defining the operators 
          \begin{eqnarray}
t_m = q^{\frac{1}{2}m_1m_2} v_{ra}^{m_1} z^{m_2} \qquad m = (m_1 , m_2) \in {{\mathbb{N}}^*}^2
          \end{eqnarray} 
we easily obtain the following result. 

\noindent {\it Proposition 11}. The commutator of the operators $t_m$ and $t_n$ reads
          \begin{eqnarray}
[t_m , t_n]_- = 2 {i} \sin \left( \frac{\pi}{2j+1} m \wedge n \right) t_{m+n}
          \end{eqnarray} 
where 
          \begin{eqnarray}
m \wedge n = m_1 n_2 - m_2 n_1 \qquad m + n = (m_1 + n_1 , m_2 + n_2).
          \end{eqnarray} 
Therefore, the linear operators $t_m$ span an infinite-dimensional Lie 
algebra.

The so-obtained Lie algebra is isomorphic to the algebra introduced in 
\cite{FFZ}. The latter result 
parallels the ones derived, on the one hand, from a study of $k$-fermions and 
of the Dirac quantum phase operator through a $q$-deformation of the harmonic 
oscillator \cite{DaoHasKib} and, on the other hand, from an investigation 
of correlation measure for finite quantum systems \cite{Ellinas}.

\subsection{Mutually unbiased bases}
We now briefly establish contact with quantum information. For this purpose, 
let us introduce the notation
          \begin{eqnarray}
k := j - m \qquad | k \rangle := | j , m \rangle \qquad d := 2j+1.
          \end{eqnarray}  
Thus, the angular momentum basis 
$\{ |j , j \rangle, |j , j-1 \rangle, \cdots, |j , -j \rangle \}$
of the finite-dimensional Hilbert space ${\cal E} (2j+1)$  reads  
$\{ | 0 \rangle, | 1 \rangle, \cdots, | d-1 \rangle \}$. Let us note 
          \begin{eqnarray}
B_d := \{ | k \rangle: k = 0, 1, \cdots, d-1 \}
          \end{eqnarray} 
the latter orthonormal basis, known as the computational basis in quantum 
information and quantum computation. From now on, the real number $a$ occurring 
in (\ref{j alpha r a in terms of jm}) shall be restricted to take the values 
$a = 0, 1, \cdots, d-1$.  

From equation (\ref{j alpha r a in terms of jm}), we can write the eigenvectors 
$| a \alpha \rangle := | j \alpha ; 0 a \rangle$ of the operator $v_{0a}$ as 
          \begin{eqnarray}
| a \alpha \rangle = \frac{1}{\sqrt{d}} \sum_{k = 0}^{d-1} 
q^{(d - k - 1)(k + 1)a / 2 - (k + 1) \alpha} | k \rangle
          \label{a alpha in terms of k}
          \end{eqnarray} 
where, for fixed $a$ ($a = 0, 1, \cdots, d-1$), the index $\alpha$ takes the
values $0, 1, \cdots, d-1$. Note that 
          \begin{eqnarray}
B_{0a} := \{ | a \alpha \rangle: \alpha = 0, 1, \cdots, d-1 \}
          \end{eqnarray} 
is another orthonormal basis of ${\cal E} (d)$. 

Proposition 8 can be transcribed in matrix form by using the 
generators $E_{x,y}$ of GL$(d , \mathbb{C})$ (see also \cite{AlbouyKibler} 
where a different normalization is used). The $d \times d$ matrix $E_{x,y}$ 
(with $x,y \in \mathbb{Z}_d$) is defined by its matrix elements
          \begin{eqnarray}
\left( E_{x,y} \right)_{kl} = \delta_{k,x} \> \delta_{l,y} \qquad 
k,l \in \mathbb{Z}_d. 
          \end{eqnarray} 
Therefore, the matrix $V_{0a}$ of the operator $v_{0a}$ in the computational basis 
$B_d$ is 
          \begin{eqnarray}
    V_{0a} = E_{d-1,0} + \sum_{k=0}^{d-2} q^{(k+1)a} E_{k,k+1}.
          \end{eqnarray} 
The eigenvectors $\varphi(a \alpha)$ 
of the matrix $V_{0a}$ are expressible in terms of the $d \times 1$ column vectors 
$e_x$ (with $x \in \mathbb{Z}_d$) defined via 
     	  \begin{eqnarray}
\left( e_x \right)_{k0} = \delta_{k , x} \qquad k \in \mathbb{Z}_d.
        \end{eqnarray}
In fact, we can check that 
        \begin{eqnarray}
\varphi(a \alpha) = \frac{1}{\sqrt{d}}
\sum_{k=0}^{d-1} q^{ (d-k-1)(k+1)a / 2 - (k+1) \alpha } e_{k} 
        \label{vector a alpha in terms of e}
        \end{eqnarray}
satisfies the eigenvalue equation
	  \begin{eqnarray}
V_{0a} \varphi(a \alpha) = q^{(d-1)a / 2 - \alpha} \varphi(a \alpha).
        \label{eigenvalue equation}
        \end{eqnarray}
Furthermore, the $d \times d$ matrix 
          \begin{eqnarray}
H_a := \sum_{\alpha = 0}^{d-1} \sum_{k=0}^{d-1} 
q^{ (d-k-1)(k+1)a /2 - (k+1) \alpha } E_{k , \alpha}
          \label{generalized Hadamard matrix}
          \end{eqnarray}
reduces the endomorphism associated with $V_{0a}$. In other words, we have
          \begin{eqnarray}
H_a ^{\dagger} V_{0a} H_a = q^{ (d-1)a / 2 } \> d \> \sum_{\alpha=0}^{d-1} 
q^{- \alpha}  E_{\alpha , \alpha}.
          \label{endomorphism}
          \end{eqnarray}
Note that $H_a$ is a  generalized Hadamard matrix in the sense that 
          \begin{eqnarray}
H_a^{\dagger} H_a = dI 
          \label{Ha adjoint par Ha}
          \end{eqnarray}
and the modulus of any element of $H_a$ is unity. Observe that the Fourier matrix 
$F$ can be written as
          \begin{eqnarray}
F = (H_0 S)^{\dagger} \qquad 
S:= \frac{1}{\sqrt{d}} \sum_{\beta = 0}^{d-1} E_{\beta, d-\beta}
          \label{connexion H0-F}
          \end{eqnarray}
where $S$ acts as a permutation matrix normalized by $\frac{1}{\sqrt{d}}$. 

As an application of (\ref{a alpha in terms of k}) or 
(\ref{vector a alpha in terms of e}) to mutually unbiased 
bases, we have the following result (see also \cite{AlbouyKibler, KiblerPlanat}). 

\noindent {\it Proposition 12}. In the case where $d = p$ is a prime integer, 
the bases $B_{0a}$ for $a = 0, 1, \cdots, p-1$ together with the computational 
basis $B_d$ constitute a complete set of $p+1$ mutually unbiased bases.

\noindent {\it Proof}. According to the definition of mutually unbiased bases 
\cite{DelIvaWotFie}, we need to prove that
          \begin{eqnarray}
| \langle k | a \alpha \rangle | = \frac{1}{\sqrt{p}}
          \label{mub1}
          \end{eqnarray} 
and 
          \begin{eqnarray}
| \langle a \alpha | b \beta \rangle | = 
\delta_{\alpha , \beta} \delta_{a , b} + \frac{1}{\sqrt{p}} (1 - \delta_{a , b}) 
          \label{mub2}
          \end{eqnarray} 
for any value of $a$, $b$, $\alpha$, $\beta$ and $k$ in $\mathbb{Z}_d$. Equation
(\ref{mub1}) simply follows from (\ref{a alpha in terms of k}) 
and equation (\ref{mub2}) was proved in \cite{AlbouyKibler} 
by making use of generalized quadratic Gauss sums. $\Box$

The interest of (\ref{a alpha in terms of k}) or (\ref{vector a alpha in terms of e}) 
with $d = p$, $p$ prime (including the case $p=2$), is that the $p^2$ vectors corresponding to the $p$ mutually 
unbiased bases besides the computational basis are obtainable from one single formula 
that is easily codable on a computer (the single formula corresponds to the diagonalization 
of only one matrix, namely, the matrix $V_{0a}$ where $a$ can take the values 
$a = 0, 1, \cdots, p-1$). In matrix form, the $p$ mutually unbiased bases besides the 
computational basis are given by the columns of the Hadamard matrices matrices $H_a$ 
($a = 0, 1, \cdots, p-1$).

Going back to $d$ arbitrary, we can check that the bases $B_{00}$, 
$B_{01}$ and $B_d$ constitute a set of 3 mutually unbiased bases. Therefore, we recover 
a well-known result according to which there exists a minimum of 3 mutually unbiased 
bases when $d$ is not a prime power. 

\section{Weyl pairs and unitary group} 
In this section, we shall focus our attention on one of the $u(d)$ 
subalgebras of $p_d$. Such a subalgebra can be constructed from a 
remarkable subset of $\{ w_{abc} : a, b, c \in \mathbb{Z}_d \}$. 
This subset is made of generalized Pauli operators. It is 
generated by the Weyl pair $(x , z)$ or $(X , Z)$ in 
matrix form. 

\subsection{Generalized Pauli operators}
Following the work by Patera and Zassenhaus \cite{PateraZassenhaus}, let us 
define the operators 
     \begin{eqnarray}
u_{ab} := w_{0ab} = x^a z^b \qquad a, b \in \mathbb{Z}_d. 
     \label{def des uab}
     \end{eqnarray}
The operators $u_{ab}$ are unitary. Note that the matrices $X^a Z^b$ of 
the operators $u_{ab}$ in the basis $b_s$ belong to the unitary group 
$U(d)$ for $d$ even or to the special unitary group $SU(d)$ for $d$ odd. 
The $d^2$ operators $u_{ab}$ shall be refered to as generalized Pauli 
operators in dimension $d$. It should be mentioned that matrices 
corresponding to the operators of type (\ref{def des uab}) were first 
introduced long time ago by Sylvester \cite{Sylvester} in order to solve 
the matrix equation $PX = XQ$; in addition, 
such matrices were used by Morris \cite{Morris} to define generalized
Clifford algebras in connection with quaternion algebras and division rings. 
The operators $u_{ab}$ satisfy the two following properties which are direct 
consequences of (\ref{trace des w(abc)}) and (\ref{commutateur des w(abc)}).

\noindent {\it Proposition 13}. The set $\{ u_{ab} : a, b \in \mathbb{Z}_d \}$ 
is an orthogonal set with respect to the 
Hilbert-Schmidt inner product. More precisely 
          \begin{eqnarray}
 {\rm Tr}_{{\cal E}(2j+1)} \left( u_{ab}^{\dagger} u_{a'b'} \right) = 
 d \>
 \delta_{a,a'} \> 
 \delta_{b,b'} 
          \label{trace de uu}
          \end{eqnarray}
where the trace has to be taken on the $d$-dimensional space ${\cal E}(2j+1)$ with $d := 2j+1$.

\noindent {\it Proposition 14}. The commutator 
                $[u_{ab} , u_{a'b'}]_-$ and the 
anti-commutator $[u_{ab} , u_{a'b'}]_+$ of $u_{ab}$ and $u_{a'b'}$ are given by 
          \begin{eqnarray}
[u_{ab} , u_{a'b'}]_{\mp} = \left( q^{-ba'} \mp q^{-ab'} \right) u_{a'' b''} 
\qquad a'' := a+a' 
\qquad b'' := b+b'. 
          \label{com anti-com}
          \end{eqnarray}
Consequently, $[u_{ab} , u_{a'b'}]_{-} = 0$ if and only if $ab' - ba' = 0$
(mod $d$) and $[u_{ab} , u_{a'b'}]_{+} = 0$ if and only if $ab' - ba' = (1/2) d$
(mod $d$). Therefore, all anti-commutators $[u_{ab} , u_{a'b'}]_{+}$ are
different from 0 if $d$ is an odd integer.  

The $d^2$ pairwise orthogonal operators $u_{ab}$ can be used as a basis of the Hilbert space 
$\mathbb{C}^{d^2}$ (with the Hilbert-Schmidt scalar product) of the operators acting on 
the Hilbert space 
$\mathbb{C}^{d}$   (with the usual scalar product). In matrix form, they give 
generalized Pauli matrices in $(2 j + 1) \times (2 j + 1)$ dimensions, the 
spin angular momentum $j = 1/2$ corresponding to the ordinary Pauli matrices. 

\noindent {\it Example 1}: $j = 1/2 \Rightarrow q = -1$ and $d = 2$. The matrices of the four 
operators $u_{ab}$ with 
$a, b = 0,1$ are 
     \begin{eqnarray}
I = X^0 Z^0 = 
\pmatrix{
  1     &0   \cr
  0     &1   \cr
} \qquad 
X = X^1 Z^0 = 
\pmatrix{
  0     &1   \cr
  1     &0   \cr
}
     \end{eqnarray}
     \begin{eqnarray}
Z = X^0 Z^1 = 
\pmatrix{
  1     &0   \cr
  0     &-1  \cr
} \qquad 
Y := X^1 Z^1 = 
\pmatrix{
  0     &-1  \cr
  1     &0   \cr
}.
     \end{eqnarray}
In terms of the usual (Hermitean and unitary) Pauli matrices 
$\sigma_x$, $\sigma_y$ and $\sigma_z$, we have $X = \sigma_x$,   
$Y = - i \sigma_y$, $Z = \sigma_z$. Note that a normalization for 
the Pauli matrices different from the conventional one is also used 
in \cite{PateraZassenhaus}. The group-theoretical approaches 
developed in \cite{PateraZassenhaus} and in the present paper lead to Pauli matrices in dimension 
$2 \times 2$ that differ from the usual Pauli matrices. This is the price one has to pay in order 
to get a systematic generalization of Pauli matrices in arbitrary dimension (see also 
\cite{PateraZassenhaus, Pittenger}). It should be observed that the commutation and anti-commutation 
relations given by (\ref{com anti-com}) with $d=2$ correspond to the well-known 
commutation and anti-commutation relations for the usual Pauli matrices (transcribed 
in the normalization 
$X^1 Z^0 = \sigma_x$, 
$X^1 Z^1 = -i \sigma_y$, 
$X^0 Z^1 = \sigma_z$).

\noindent {\it Example 2}: $j = 1 \Rightarrow q = \exp(2 \pi i/3)$ and $d = 3$. The matrices 
of the nine operators $u_{ab}$ with $a, b = 0,1,2$, viz., 
     \begin{eqnarray}
X^0 Z^0 = I      \qquad 
X^1 Z^0 = X      \qquad
X^2 Z^0 = X^2    \qquad
X^0 Z^1 = Z      \qquad
X^0 Z^2 = Z^2    
     \end{eqnarray}
     \begin{eqnarray}
X^1 Z^1 = X Z    \qquad 
X^2 Z^2          \qquad
X^2 Z^1 = X^2 Z  \qquad  
X^1 Z^2 = X Z^2  
     \end{eqnarray}
are 
     \begin{eqnarray}
I = 
\pmatrix{
  1     &0     &0   \cr
  0     &1     &0   \cr
  0     &0     &1   \cr
} \qquad 
X = 
\pmatrix{
  0     &1     &0   \cr
  0     &0     &1   \cr
  1     &0     &0   \cr
} \qquad 
X^2 = 
\pmatrix{
  0     &0     &1   \cr
  1     &0     &0   \cr
  0     &1     &0   \cr
}
     \end{eqnarray}
     \begin{eqnarray}
Z = 
\pmatrix{
  1     &0     &0     \cr
  0     &q     &0     \cr
  0     &0     &q^2   \cr
} \qquad 
Z^2 = 
\pmatrix{
  1     &0       &0   \cr
  0     &q^2     &0   \cr
  0     &0       &q   \cr
} \qquad 
X Z = 
\pmatrix{
  0     &q     &0     \cr
  0     &0     &q^2   \cr
  1     &0     &0     \cr
}
     \end{eqnarray}
     \begin{eqnarray}
X^2 Z^2 = 
\pmatrix{
  0     &0       &q     \cr
  1     &0       &0     \cr
  0     &q^2     &0     \cr
} \qquad 
X^2 Z =  
\pmatrix{
  0     &0     &q^2     \cr
  1     &0     &0       \cr
  0     &q     &0       \cr
} \qquad 
X Z^2 =  
\pmatrix{
  0     &q^2     &0     \cr
  0     &0       &q     \cr
  1     &0       &0     \cr
}.
     \end{eqnarray}
These matrices differ from the Gell-Mann matrices \cite{Gell-MannNe'emanOkubo} used in elementary particle 
physics. They constitute a natural extension of the Pauli matrices in dimension 
$3 \times 3$ (see also \cite{PateraZassenhaus, Pittenger}).

\subsection{The unitary group in the generalized Pauli basis}
From Proposition 14, it is clear that the set 
$\{ u_{ab} : a,b = 0, 1, \cdots, d-1 \}$ can be used as a set of generators of 
the Lie group $U(d)$. Thus the generalized Pauli matrices $X$ 
and $Z$ form an integrity basis for the Lie algebra of $U(d)$. This can be 
precised by the two propositions below.

\noindent {\it Proposition 15}. The set $\{ X^a Z^b : a,b = 0, 1, \cdots, d-1 \}$ 
form a basis for the Lie algebra $u(d)$ of the unitary group $U(d)$ for $d$ 
arbitrary. The Lie brackets of $u(d)$ in such a basis (that we denote as the 
Pauli basis) are given by
     \begin{eqnarray}
[ X^a Z^b , X^{a'} Z^{b'} ]_- = \sum_{a''b''} (ab,a'b';a''b'') X^{a''} Z^{b''} 
     \label{Lie brackets en XZ}
     \end{eqnarray}
where the structure constants $(ab,a'b';a''b'')$ read 
     \begin{eqnarray}
(ab,a'b';a''b'') = \delta(a'', a+a') 
                     \delta(b'', b+b') \left( q^{- ba'} - q^{- ab'} \right) 
     \label{constantes de structure en XZ}
     \end{eqnarray}
with $a, b, a', b' = 0, 1, \cdots, d-1$ (mod $d$). The structure constants 
$(ab,a'b';a''b'')$ with $a''=a+a'$ and $b''=b+b'$ are cyclotomic polynomials 
associated with $d$. They vanish for $ab' - ba' = 0$ (mod $d$).  

\noindent {\it Proposition 16}. In the case where $d=p$ is a prime integer, the 
Lie algebra $su(p)$ of the special unitary group $SU(p)$ can be decomposed into 
a direct sum of $p+1$ abelian subalgebras of dimension $p-1$. More precisely 
     \begin{eqnarray}
su(p) = 
{ v}_0     \oplus 
{ v}_1     \oplus 
\cdots      \oplus      
{ v}_{p}   
     \label{decompo de su(p)}   
     \end{eqnarray}
where each of the $p+1$ subalgebras ${ v}_0, { v}_1, \cdots, { v}_p$ is a Cartan 
subalgebra generated by a set of $p - 1$ commuting matrices. The various sets are                 
           \begin{eqnarray}  
{\cal V}_1       &:=      &  \{ X^1 Z^0    , X^2 Z^0    , \cdots, X^{p-1}     Z^0 \}   
    \label{ensemble V1}
  \\
{\cal V}_2       &:=     &  \{ X^1 Z^1    , X^2 Z^2    , \cdots, X^{p-1} Z^{p-1}  \}  
  \\
{\cal V}_3       &:=     &  \{ X^1 Z^2    , X^2 Z^4    , \cdots, X^{p-1} Z^{p-2}  \} 
  \\
                 &\vdots & 
  \\
 {\cal V}_{p-1}   &:=      &  \{ X^1 Z^{p-2}, X^2 Z^{p-4},  \cdots, X^{p-1}     Z^2  \} 
  \\  
 {\cal V}_{p}   &:=      &  \{ X^1 Z^{p-1}, X^2 Z^{p-2},  \cdots, X^{p-1}     Z^1  \} 
          \end{eqnarray} 
and
          \begin{eqnarray} 
{\cal V}_0         &:=      &  \{ X^0 Z^1    , X^0 Z^2    , \cdots, X^0     Z^{p-1}  \} 
    \label{ensemble V0}
          \end{eqnarray} 
for ${ v}_1, { v}_2, \cdots, { v}_{p}$ and ${ v}_0$, respectively. 

\noindent {\it Proof}. The proof of Proposition 15 is straightforward: It 
follows from (\ref{trace de uu}) and (\ref{com anti-com}). For Proposition 16, 
we need to pass from $u(p)$ to its subalgebra $su(p)$. A basis for the Lie 
algebra $su(p)$ of $SU(p)$ is provided with the set 
$\{ X^a Z^b : a, b = 0, 1, \cdots, p-1 \} \setminus \{ X^0 Z^0 \}$. Then, in 
order to prove Proposition 16, it suffices to verify that the $p+1$ sets 
(or classes) 
${\cal V}_0$, 
${\cal V}_1$, 
$\cdots$,     
${\cal V}_{p-1}$, 
and
${\cal Z}$ 
constitute a partition of 
$\{ X^a Z^b : a, b = 0, 1, \cdots, p-1 \} \setminus \{ X^0 Z^0 \}$ 
and that the $p-1$ operators in each set commute one with each other.  
Proposition 16 takes its origin in a remark \cite{Kib polar decomp}
according to which the rank of $su(p)$ is $p-1$ 
so that the case of $p+1$ sets containing $p-1$ commuting operators occurs as a 
limiting case. The decomposition (\ref{decompo de su(p)}), also valid for $sl(p , \mathbb{C})$, 
was first derived in \cite{PateraZassenhaus} in connection with the determination of finest 
gradings of Lie algebras of type $A_{p-1}$. It is little known that a decomposition of type 
(\ref{decompo de su(p)}) was conjectured almost three decades ago \cite{KKU}
for the more general case where $p$ is replaced by a prime power (see also 
\cite{autres decomp of sun}). $\Box$ 

\noindent {\it Example 3}. For the purpose of clarifying the production process 
of the sets ${\cal V}_{i}$ (for $i = 0, 1, \cdots, p$), let us consider 
the case $p = 7 \Leftrightarrow j = 3$). Equations (\ref{ensemble V1})-(\ref{ensemble V0}) give
           \begin{eqnarray}  	   
{\cal V}_0       &=      &  \{ ( 01 ), ( 02 ), ( 03 ), ( 04 ), ( 05 ), ( 06 ) \} 
  \\	   	   
{\cal V}_1       &=      &  \{ ( 10 ), ( 20 ), ( 30 ), ( 40 ), ( 50 ), ( 60 ) \} 
  \\
{\cal V}_2      &=      &  \{ ( 11 ), ( 22 ), ( 33 ), ( 44 ), ( 55 ), ( 66 ) \} 
  \\
{\cal V}_3      &=      &  \{ ( 12 ), ( 24 ), ( 36 ), ( 41 ), ( 53 ), ( 65 ) \} 
  \\
{\cal V}_4      &=      &  \{ ( 13 ), ( 26 ), ( 32 ), ( 45 ), ( 51 ), ( 64 ) \} 
  \\
{\cal V}_5      &=      &  \{ ( 14 ), ( 21 ), ( 35 ), ( 42 ), ( 56 ), ( 63 ) \} 
  \\
{\cal V}_6      &=      &  \{ ( 15 ), ( 23 ), ( 31 ), ( 46 ), ( 54 ), ( 62 ) \} 
  \\
{\cal V}_7      &=      &  \{ ( 16 ), ( 25 ), ( 34 ), ( 43 ), ( 52 ), ( 61 ) \} 
          \end{eqnarray} 
where $(ab)$ is used as an abbreviation of $X^a Z^b$.

At this stage, it should be stressed that decompositions of type 
(\ref{decompo de su(p)}-\ref{ensemble V0}) are especially useful for the 
construction of mutually unbiased bases \cite{AlbouyKibler, DelIvaWotFie}. 
Along this vein, the common eigenvectors of each of the $p+1$ subalgebras 
${ v}_0, { v}_1, \cdots, { v}_p$ give rise to $p+1$ mutually unbiased bases. 
Unfortunately, finding a general formula for the Lie brackets of each pair 
of the Cartan subalgebras is a difficult problem for which we have no answer.

\noindent {\it Counterexample 1}. For $d=4  \Leftrightarrow j = 3/2$ 
($\Rightarrow a, b = 0,1, 2, 3$), Proposition 15 is valid but Proposition 
16 does not apply. Indeed, the 16 unitary operators $u_{ab}$ corresponding to 
\begin{eqnarray} 
ab = 01, 02, 03, 10, 20, 30, 11, 22, 33, 12, 13, 21, 23, 31, 32, 00
\label{les ab en dim 4}
\end{eqnarray}
are linearly independent and span the Lie algebra of $U(4)$ but they give only 3
disjoint sets, viz., $\{ (01), (02), (03) \}$, $\{ (10), (20), (30) \}$ and  
$\{ (11), (22), (33) \}$, containing each 3 commuting operators, where 
here again $(ab)$ stands for $X^a Z^b$. However, it is not possible to
partition the set (\ref{les ab en dim 4}) in order to get a decomposition
similar to (\ref{decompo de su(p)}). Nevertheless, it is possible to find 
another basis of $u(4)$ which can be partioned in a way yielding a
decompostion similar to (\ref{decompo de su(p)}). This can be achieved by 
working with tensorial products of the matrices $X^a Z^b$ corresponding 
to $p=2$. In this respect, let us consider the product $u_{a_1b_1} \otimes u_{a_2b_2}$, 
where $u_{a_ib_i}$ with $i = 1,2$ are Pauli operators for $p=2$. Then, by using
the abbreviation $(a_1b_1a_2b_2)$ for $u_{a_1b_1} \otimes u_{a_2b_2}$ or 
$X^{a_1} Z^{b_1} \otimes X^{a_2} Z^{b_2}$, it can be checked that the 5 disjoint 
sets
\begin{eqnarray} 
\{ (1011), (1101), (0110) \} 
\label{set1} \\
\{ (1001), (0111), (1110) \} 
\label{set2} \\
\{ (1010), (1000), (0010) \} 
\label{set3} \\
\{ (1111), (1100), (0011) \} 
\label{set4} \\
\{ (0101), (0100), (0001) \} 
\label{set5}
\end{eqnarray} 
consist each of 3 commuting unitary operators and that the Lie algebra $su(4)$ 
is spanned by the union of the 5 sets. It is to be emphasized that the 15 
operators (\ref{set1}-\ref{set5}) are underlaid by the geometry of the generalized 
quadrangle of order 2 \cite{PlanatGPM11}. In this geometry, the five sets given 
by (\ref{set1}-\ref{set5}) correspond to a spread of this quadrangle, i.e., to a 
set of 5 pairwise skew lines \cite{PlanatGPM11}.

The considerations of Counterexample 1 can be generalized in the case 
$d := d_1 d_2 \cdots d_e$, $e$ being an integer greater or equal to
$2$. Let us define 
\begin{eqnarray} 
u_{AB} := u_{a_1b_1} \otimes u_{a_2b_2} \otimes \cdots \otimes u_{a_eb_e}
\qquad A := a_1, a_2, \cdots, a_e
\qquad B := b_1, b_2, \cdots, b_e
\label{u_AB}
\end{eqnarray} 
where $u_{a_1b_1}, u_{a_2b_2}, \cdots,  u_{a_eb_e}$ are generalized Pauli operators 
corresponding to the dimensions $d_1, d_2, \cdots, d_e$ respectively. (The operators 
$u_{AB}$ are elements of the group $P_{d_1} \otimes P_{d_2} \otimes \cdots \otimes 
P_{d_e}$. We follow \cite{PateraZassenhaus} by calling the operators $u_{AB}$ 
generalized Dirac operators since the ordinary Dirac operators correspond to $P_{2} 
\otimes P_{2}$.) In addition, let $q_1, q_2, \cdots, q_e$ be the $q$-factor associated 
with $d_1, d_2, \cdots, d_e$ respectively ($q_j := \exp (2 \pi i / d_j)$). Then, 
Propositions 13, 14 and 15 can be generalized as follows.

\noindent {\it Proposition 17}. The operators $u_{AB}$ are unitary and satisfy the 
orthogonality relation
          \begin{eqnarray}
 {\rm Tr}_{{\cal E}(d_1 d_2 \cdots d_e)} \left( u_{AB}^{\dagger} u_{A'B'} \right) = 
 d_1 d_2 \cdots d_e \>
 \delta_{A,A'} \> 
 \delta_{B,B'} 
          \label{trace de uABuA'B'}
          \end{eqnarray}
where 
         \begin{eqnarray}
\delta_{A,A'} := \delta_{a_1,a_1'} \delta_{a_2,a_2'} \cdots \delta_{a_e,a_e'} \qquad 
\delta_{B,B'} := \delta_{b_1,b_1'} \delta_{b_2,b_2'} \cdots \delta_{b_e,b_e'}. 
         \label{delta de AA' et BB'}
          \end{eqnarray}
The commutator 
                $[u_{AB} , u_{A'B'}]_-$ and the 
anti-commutator $[u_{AB} , u_{A'B'}]_+$ of $u_{AB}$ and $u_{A'B'}$ are given by 
          \begin{eqnarray}
[u_{AB} , u_{A'B'}]_{\mp} = \left( \prod_{j=1}^e q_j^{-b_ja_j'} \mp 
                                   \prod_{j=1}^e q_j^{-a_jb_j'} \right) u_{A'' B''}  
          \label{com anti-com en AB}
          \end{eqnarray}
with
          \begin{eqnarray}
A'' := a_1 + a_1', a_2 + a_2', \cdots, a_e + a_e' \qquad  
B'' := b_1 + b_1', b_2 + b_2', \cdots, b_e + b_e'.   
          \label{A''B''}
          \end{eqnarray}
The set 
$\{ u_{AB} : A, B \in \mathbb{Z}_{d_1} \otimes \mathbb{Z}_{d_2} \otimes \cdots \otimes \mathbb{Z}_{d_e} \}$ 
of the $d_1^2 d_2^2 \cdots d_e^2$ 
unitary operators $u_{AB}$ form a basis for the Lie algebra $u(d_1 d_2 \cdots d_e)$ 
of the group $U(d_1 d_2 \cdots d_e)$. 
In the special case where $d_1 = d_2 = \cdots = d_e = p$ with $p$ a prime integer 
(or equivalently $d = p^e$), we have $[u_{AB} , u_{A'B'}]_{-} = 0$ if and only if 
          \begin{eqnarray}
\sum_{j=1}^e a_jb_j' - b_ja_j' = 0 \quad ({\rm mod} \ p) 
          \end{eqnarray}
and $[u_{AB} , u_{A'B'}]_{+} = 0$ if and only if 
          \begin{eqnarray}
\sum_{j=1}^e a_jb_j' - b_ja_j' = \frac{1}{2} p \quad ({\rm mod} \ p) 
          \label{anti-com}
          \end{eqnarray}
so that there are vanishing anti-commutators only if $p = 2$. For $d = p^e$, there exists a 
decomposition of the set $\{u_{AB} : A, B \in \mathbb{Z}_{p}^{\otimes e}\} \setminus \{I\}$ 
that corresponds to a decomposition of the Lie algebra $su(p^e)$ into $p^e +1$ abelian 
subalgebras of dimension $p^e - 1$. 

\noindent {\it Proof}. The proof of (\ref{trace de uABuA'B'})-(\ref{anti-com}) is based on 
repeated application of Proposition 13. For $d=p^e$, we know from 
\cite{BandyoGPM5, LawrenceGPM6, Lawrence2} that the set $\{u_{AB} : A, B \in \mathbb{Z}_{p}^{\otimes e}\} \setminus \{I\}$ 
(consisting of $p^{2e} -1$ unitary operators that are pairwise orthogonal), 
which provides a basis for $su(p^e)$, can be partioned into $p^e + 1$
disjoint classes containing each $p^{e} -1$ commuting operators. Therefore, 
there exists a decompostion of $su(p^e)$ into a direct sum of $p^e + 1$ 
subalgebras of dimension $p^{e} -1$. (There is a one-to-one correspondence between 
the $p^e + 1$ subalgebras and the $p^e + 1$ mutually unbiased bases in $\mathbb{C}^{p^e}$.) $\Box$

\section{Closing remarks}

Starting from an abstract definition of the Heisenberg-Weyl group, combined with 
a polar decompostion of $SU(2)$ arising from  angular  momentum  theory, we have 
analysed in a detailed way the interelationship between Weyl pairs,  generalized 
Pauli operators and generalized Pauli group.  The interest of these developments 
for the unitary group $U(d)$, $d$ arbitrary, have been underlined with a special 
emphasis for a decomposition of  $su(d)$  when  $d$  is the power of a prime. We 
would like to close with two remarks. 

In arbitrary dimension $d$, the number of mutually unbiased bases in $\mathbb{C}^d$ 
is less or equal to $d + 1$ \cite{BandyoGPM5, DelIvaWotFie}.  Proposition 17 suggests 
the  following  remark.    To prove that the  number  of mutually unbiased bases in 
$\mathbb{C}^d$ is $d + 1$ for $d$ arbitrary amounts to prove that it is possible to 
find  a  decomposition  of  the  Lie  algebra $su(d)$  into the direct sum of $d+1$ 
abelian subalgebras of dimension $d-1$.   Therefore, if such a decomposition cannot 
be  found,  it  would  result  that  the  number  of  mutually  unbiased  bases  in 
$\mathbb{C}^d$  is  less  than  $d + 1$  when  $d$  is  not a prime power 
(cf. Conjectures 5.4 and 5.5 by Boykin {\it et al} \cite{autres decomp of sun}). 

The Pauli group or discrete Heisenberg-Weyl group $P_d \equiv HW(\mathbb{Z}_d)$ 
plays an important role in deriving mutually unbiased bases in finite-dimensional 
Hilbert spaces. We know that the concept of mutually unbiased bases also exists 
in infinite dimension \cite{MUBs continues}. In this connection, the infinite or 
ordinary Heisenberg-Weyl group $HW(\mathbb{R})$ might be of interest for 
constructing mutually unbiased bases in infinite-dimensional Hilbert spaces.

\section*{Acknowledgments}

The author is indebted to Olivier Albouy, Anne-C\'eline Baboin, Hans Havlicek, 
Michel Planat and Metod Saniga for interesting discussions. He wishes to thank 
Arthur Pittenger, Michel Planat, Metod Saniga, Apostol Vourdas and Bernardo 
Wolf as well as the Referees for bringing his attention to additional 
publications and for useful comments and suggestions on the manuscript. This 
work started while the author enjoyed the scientific atmosphere 
of the workshop `Finite projective geometries in quantum theory' held in 
Tatranska-Lomnica in August 2007. Financial support from the ECO-NET project 
12651NJ `Geometries over finite rings and the properties of mutually unbiased 
bases' is gratefully acknowledged.

\baselineskip = 0.50 true cm

\end{document}